\documentclass[aps,prd,amsmath,preprint,noshowpacs,nofootinbib,a4paper]{revtex4}

\allowdisplaybreaks[1]
\begin{document}
\fontsize{12pt}{13.5pt}\selectfont

\title{Mixing of fermion fields of opposite parities and baryon resonances}
\author{A.E. Kaloshin}
\email{kaloshin@isu.ru}
\author{E.A. Kobeleva}
\email{elenyich@mail.ru}
\affiliation{Irkutsk State University, K. Marx str., 1, 664003, Irkutsk, Russia}
\author{V.P. Lomov}
\email{lomov.vl@icc.ru} \affiliation{Institute for System Dynamics
and Control Theory, RAS, Lermontov str., 134, 664043, Irkutsk,
Russia}
\begin{abstract}
  We consider a loop mixing of two fermion fields of opposite parities whereas
  the parity is conserved in a Lagrangian. Such kind of mixing is specific for
  fermions and has no analogy in boson case.  Possible applications of this
  effect may be related with physics of baryon resonances. The obtained matrix
  propagator defines a pair of unitary partial amplitudes which describe the
  production of resonances of spin $J$ and different parity ${1/2}^{\pm}$ or
  ${3/2}^{\pm}$.
  \\
  The use of our amplitudes for joint description of $\pi N$ partial waves
  $P_{13}$ and $D_{13}$ shows that the discussed effect is clearly seen in these
  partial waves as the specific form of interference between resonance and
  background. Another interesting application of this effect may be a pair of
  partial waves $S_{11}$ and $P_{11}$ where the picture is more complicated due
  to presence of several resonance states.
\end{abstract}
\pacs{11.80Et, 14.20Gk, 11.80Jy.}

\maketitle

\section{Introduction}
\label{sec:intro}

Mixing of states (fields) is a well-known phenome- non existing in the systems
of neutrinos \cite{Pon58}, quarks \cite{Cab63} and hadrons. In hadron systems
the mixing effects are essential not only for $K^0$- and $D^0$-mesons but also
for the broad overlapping resonances. As for theoretical description of mixing
phenomena, a general tendency with time and development of experiment consists
in transition from a simplified quantum-mechanical description to the quantum
field theory methods (see e.g. review \cite{Beu03}, more recent papers
\cite{Esp02,Bla04,Mac05,Kni06,Dur08} and references therein).

Mixing of fermion fields has some specifics as compared with boson
case. Firstly, there exists $\gamma$-matrix structure in a propagator. Secondly,
fermion and antifermion have the opposite $P$-parity, so fermion propagator
contains contributions of different parities.  As a result, besides a standard
mixing of fields with the same quantum numbers, for fermions there exists a
mixing of fields with opposite parities (OPF-mixing), even if the parity is
conserved in Lagrangian.

Such a possibility for fermion mixing has been noted in \cite{Kal06}. In this
paper we study this effect in detail and apply it to the baryon resonances
production in $\pi N$ reaction.

In section 2 we consider a standard mixing of fermion fields of the same
parity. Following to \cite{Kal04,Kal06,Gon07} we use the off-shell projection
basis to solve the Dyson--Schwinger equation, it simplilies all manipulations
with $\gamma$-matrices and, moreover, clarifies the meaning of formulas. The use
of this basis leads to separation of $\gamma$-matrix structure, so in standard
case we come to studying of a mixing matrix, which is very similar to boson
mixing matrix.

In section 3 we derive a general form of matrix dressed propagator with
accounting of the OPF-mixing. In contrast to standard case the obtained
propagator contains $\gamma^5$ terms, even if parity is conserved in vertexes.

Section 4 is devoted to more detailed studying of considered OPF-mixing in
application to production of resonances $J^P={1/2}^\pm, I=1/2$ in $\pi N$
scattering. First estimates demonstrates that the considered mixing generates
marked effects in $\pi N$ partial waves, changing a typical resonance curve.
Comparison of the obtained multichannel hadron amplitudes with $K$-matrix
parameterization shows that our amplitudes may be considered as a specific
variant of analytical $K$-matrix.

In section 5 we consider OPF-mixing for case of two vector-spinor
Rarita-Schwinger fields $\Psi^\mu$, describing spin-$3/2$ particles, and apply
the obtained hadron amplitudes for descriptions of $\pi N$ partial waves
$P_{13}$ and $D_{13}$.

Conclusion contains discussion of results.

In Application there are collected some details of calculations, concerning the
production of spin-$3/2$ resonances.

\section{Mixing of fermion fields of the same parity}
\label{sec:same_parity}

Let us start from the standard picture when the mixing fermions have the same
quantum numbers. To obtain the dressed fermion propagator $G(p)$ one should
perform the Dyson summation or, equivalently, to solve the Dyson--Schwinger
equation:
\begin{equation}\label{lab1}
  G(p)=G_{0}+G\Sigma G_{0},
\end{equation}
where $G_{0}$ is a free propagator and $\Sigma$ is a self-energy:
\begin{equation}
  \Sigma(p)=A({p}^{2})+\widehat{p}B({p}^{2}).
\end{equation}

We will use the off-shell projection operators $\Lambda^{\pm}$:
\begin{equation*}
  \Lambda^{\pm}=\frac{1}{2} \Big(1\pm\frac{\hat{p}}{W} \Big),
\end{equation*}
where $W=\sqrt{p^2}$ is energy in the rest frame.

Main properties of projection operators are:
\begin{equation*}
  \Lambda^{\pm}\Lambda^{\pm}=\Lambda^{\pm},\quad
  \Lambda^{\pm}\Lambda^{\mp}=0,\quad
  \Lambda^{\pm}\gamma^{5}=\gamma^{5}\Lambda^{\mp},
\end{equation*}
\begin{equation*}
  \Lambda^{+}+\Lambda^{-}=1,\quad
  \Lambda^{+}-\Lambda^{-}=\frac{\hat{p}}{W}.
\end{equation*}
Let us rewrite the equation \eqref{lab1} expanding all elements in the basis of
projection operators:
\begin{equation}
  G=\sum_{M=1}^{2}\mathcal{P}_{M}G^{M},
\end{equation}
where we have introduced the notations:
\begin{equation*}
  \mathcal{P}_{1}=\Lambda^{+},\quad
  \mathcal{P}_{2}=\Lambda^{-}.\quad
\end{equation*}

In this basis the Dyson--Schwinger equation is reduced to equations on scalar
functions:
\begin{equation} \label{lab3}
  G^{M}=G^{M}_{0}+ G^{M} \Sigma^{M} G^{M}_{0}, \qquad M=1,2 ,
\end{equation}
or
\begin{equation} \label{lab4}
  \Big(G^{-1}\Big){\vphantom{G^{-1}}}^{M} =
  \Big(G^{-1}_0\Big){\vphantom{G^{-1}}}^{M} - \Sigma^M . \
\end{equation}

The solution of \eqref{lab3} for dressed propagator looks like:
\begin{equation}
\begin{split}
  \Big(G^{-1}\Big){\vphantom{G^{-1}}}^{1} &=
  \Big(G^{-1}_0\Big){\vphantom{G^{-1}}}^{1} - \Sigma^{1} = W-m-A(W^2)-W B(W^2),\\
  \Big(G^{-1}\Big){\vphantom{G^{-1}}}^{2} &=\Big(G^{-1}_0\Big){\vphantom{G^{-1}}}^{2} - \Sigma^{2}
   = -W-m-A(W^2)+W B(W^2)
\end{split}
\end{equation}
where $A$, $B$ are commonly used components of the self-energy. The coefficients
in the projection basis have the obvious property:
\begin{equation*}
  \Sigma^{2}(W)=\Sigma^{1}(-W).
\end{equation*}

When we have two fermion fields $\Psi_{i}$ , the including of interaction leads
also to mixing of these fields. In this case the Dyson--Schwinger equation
\eqref{lab1} acquire matrix indices:
\begin{equation}
  G_{ij}=(G_{0})_{ij} + G_{ik}\Sigma_{kl}(G_{0})_{lj}, \quad i,j,k,l=1,2.
\end{equation}

Therefore one can use the same equation \eqref{lab1} assuming all coefficients
to be matrices.

The simplest variant is when the fermion fields $\Psi_{i}$ have the same quantum
numbers and the parity is conserved in the Lagrangian. In this case the inverse
propagator following \eqref{lab4} has the form:
\begin{equation}
  \begin{split}
    G^{-1}&=\mathcal{P}_{1}S^{1}(W)+\mathcal{P}_{2}S^{2}(W) = \\[3mm]
         & = \mathcal{P}_{1}
      \begin{pmatrix}
        W-m_1-\Sigma^{1}_{11} & -\Sigma^{1}_{12}\\
        -\Sigma^{1}_{21}  & W-m_2-\Sigma^{1}_{22}
      \end{pmatrix}
    +\\[3mm]
    &+ \mathcal{P}_{2}
      \begin{pmatrix}
        -W-m_1-\Sigma^{2}_{11} & -\Sigma^{2}_{12}\\
        -\Sigma^{2}_{21} & -W-m_2-\Sigma^{2}_{22}
      \end{pmatrix}.
  \end{split}
\end{equation}

The matrix coefficients as before have the symmetry property
$S^{2}(W)=S^{1}(-W)$. To obtain the matrix dressed propagator $G(p)$ one should
reverse the matrix coefficients in projection basis:
\begin{equation}\label{mix1}
  \begin{split}
    G(p)&=\mathcal{P}_{1}(S^{1}(W))^{-1}+\mathcal{P}_{2}(S^{2}(W))^{-1} = \\[3mm]
    &= \mathcal{P}_{1}
      \begin{pmatrix}
        \dfrac{W-m_2-\Sigma^{1}_{22}}{\Delta_{1}} &
        -\dfrac{\Sigma^{1}_{12}}{\Delta_{1}}\\[4mm]
        -\dfrac{\Sigma^{1}_{21}}{\Delta_{1}} & \dfrac{W-m_1-\Sigma^{1}_{11}}{\Delta_{1}}
      \end{pmatrix}
    +\\[3mm]
    &+ \mathcal{P}_2
      \begin{pmatrix}
        \dfrac{-W-m_2-\Sigma^{2}_{22}}{\Delta_{2}} & -\dfrac{\Sigma^{2}_{12}}{\Delta_{2}}\\[4mm]
        -\dfrac{\Sigma^{2}_{21}}{\Delta_{2}} & \dfrac{-W-m_1-\Sigma^{2}_{11}}{\Delta_{2}}
      \end{pmatrix},
  \end{split}
\end{equation}
where
\begin{equation*}
  \begin{split}
    \Delta_{1}&=\big(W-m_1-\Sigma^{1}_{11}\big)\big(W-m_2-\Sigma^{2}_{22}\big)
               -\Sigma^{4}_{12}\Sigma^{3}_{21},\\
    \Delta_{2}&=\big(-W-m_1-\Sigma^{2}_{11}\big)\big(-W-m_2-\Sigma^{1}_{22}\big)
               -\Sigma^{3}_{12}\Sigma^{4}_{21}=\Delta_{1}\big(W\to-W\big).
  \end{split}
\end{equation*}

We see that with use of projection basis the problem of fermion mixing is
reduced to studying of the same mixing matrix as for bosons besides the obvious
replacement $s-m^{2} \rightarrow W-m$.

\section{Mixing of fermion fields of opposite P-parities}
\label{sec:diff_parity}

Let us consider the joint dressing of two fermion fields of opposite parities
provided that the parity is conserved in a vertex. In this case the diagonal
transition loops $\Sigma_{ii}$ contain only $I$ and $\hat{p}$ matrices, while
the off-diagonal ones $\Sigma_{12},\Sigma_{21}$ must contain $\gamma^{5}$.
Projection basis should be supplemented by elements containing $\gamma^{5}$, it
is convenient to choose the $\gamma$-matrix basis as:
\begin{equation}\label{PL}
  \mathcal{P}_{1}=\Lambda^{+},\quad
  \mathcal{P}_{2}=\Lambda^{-},\quad
  \mathcal{P}_{3}=\Lambda^{+}\gamma^{5},\quad
  \mathcal{P}_{4}=\Lambda^{-}\gamma^{5}.\quad
\end{equation}

In this case the $\gamma$-matrix decomposition has four terms:
\begin{equation} \label{lab5}
  S=\sum_{M=1}^{4}\mathcal{P}_{M}S^{M},
\end{equation}
where the coefficients $S^{M}$ are matrices and have the obvious symmetry
properties:
\begin{equation}
  S^2(W)=S^1(-W), \quad
  S^4(W)=S^3(-W).
\end{equation}

Inverse propagator in this basis looks as:
\begin{equation}  \label{lab6}
  \begin{split}
    S(p)&=\mathcal{P}_1
      \begin{pmatrix}
        W-m_1-\Sigma^{1}_{11} & 0\\
        0 & W-m_2-\Sigma^{1}_{22}
      \end{pmatrix}
    +\\[3mm]
    &+ \mathcal{P}_2
      \begin{pmatrix}
        -W-m_1-\Sigma^{2}_{11} & 0\\
        0 & -W-m_2-\Sigma^{2}_{22}
      \end{pmatrix}
    +\\[3mm]
    &+\mathcal{P}_3
      \begin{pmatrix}
        0 & -\Sigma^{3}_{12}\\
        -\Sigma^{3}_{21} & 0
      \end{pmatrix}
    + \mathcal{P}_4
      \begin{pmatrix}
        0 & -\Sigma^{4}_{12}\\
        -\Sigma^{4}_{21} & 0
      \end{pmatrix},
  \end{split}
\end{equation}
where the indexes $i,j=1,2$ in the self-energy $\Sigma^{M}_{ij}$ numerate
dressing fermion fields and the indexes $M={1,\dots 4}$ are refered to the
$\gamma$-matrix decomposition \eqref{lab5}.

Elements of the basis \eqref{PL} have simple multiplicative properties (see
Table~\ref{multiplic}), so reversing of \eqref{lab6} present no special problems
\cite{Kal06}.

\begin{table}
  \caption{\label{multiplic} Multiplicative properties of elements of basis \eqref{PL}.}
  \begin{center}
    \begin{tabular}{c|cccc}
      & $\mathcal{P}_1$ & $\mathcal{P}_2$ & $\mathcal{P}_3$ &
      $\mathcal{P}_4$ \\ \hline
      $\mathcal{P}_1$ & $\mathcal{P}_1$ & 0 & $\mathcal{P}_3$ & 0 \\
      $\mathcal{P}_2$ & 0 & $\mathcal{P}_2$ & 0 & $\mathcal{P}_4$ \\
      $\mathcal{P}_3$ & 0 & $\mathcal{P}_3$ & 0 & $\mathcal{P}_1$ \\
      $\mathcal{P}_4$ & $\mathcal{P}_4$ & 0 & $\mathcal{P}_2$ & 0\\
    \end{tabular}
  \end{center}
\end{table}
Reversing of \eqref{lab6} gives the matrix dressed propagator of the form:
\begin{equation}
  \label{lab7}
  \begin{split}
    G=& \mathcal{P}_1
      \begin{pmatrix}
        \dfrac{-W-m_2-\Sigma^{2}_{22}}{\Delta_1} & 0\\
        0 & \dfrac{-W-m_1-\Sigma^{2}_{11}}{\Delta_2}
      \end{pmatrix} + \\[3mm]
    +& \mathcal{P}_2
      \begin{pmatrix}
        \dfrac{W-m_2-\Sigma^{1}_{22}}{\Delta_2} & 0\\
        0 & \dfrac{W-m_1-\Sigma^{1}_{11}}{\Delta_1}
      \end{pmatrix} + \\[3mm]
    +& \mathcal{P}_3
      \begin{pmatrix}
        0 & \dfrac{\Sigma^{3}_{12}}{\Delta_1}\\
        \dfrac{\Sigma^{3}_{21}}{\Delta_2} & 0
      \end{pmatrix}+
    \mathcal{P}_4
      \begin{pmatrix}
        0 & \dfrac{\Sigma^{4}_{12}}{\Delta_2}\\
        \dfrac{\Sigma^{4}_{21}}{\Delta_1} & 0
      \end{pmatrix}.
  \end{split}
\end{equation}

Here
\begin{equation*}
  \begin{split}
    \Delta_{1}&=\big(W-m_1-\Sigma^{1}_{11}\big)\big(-W-m_2-\Sigma^{2}_{22}\big)
               -\Sigma^{3}_{12}\Sigma^{4}_{21},\\
    \Delta_{2}&=\big(-W-m_1-\Sigma^{2}_{11}\big)\big(W-m_2-\Sigma^{1}_{22}\big)
               -\Sigma^{4}_{12}\Sigma^{3}_{21}=\Delta_{1}\big(W\to-W\big).
  \end{split}
\end{equation*}

The propagator \eqref{lab7} can be compared with the standard case of mixing
(fermion fields of the same parity) \eqref{mix1}.

\section{$\pi N$ scattering and mixing of baryons $1/2^{\pm}$}
\label{sec:scattering}

As for possible applications of considered effect to description of baryon
resonances, this is, first of all, $\pi N$ scattering, where the high accuracy
data exist and detailed partial wave analysis has been performed
\cite{Cut79,Koc85,Hoh93,Arn95,Arn06}.

\subsection{Partial waves}

Let us consider an effect of OPF-mixing on the production of baryon resonances
of spin-parity $J^{P}={1/2}^{\pm}$ and isospin $I=1/2$ in $\pi N$-collisions.

Simplest effective Lagrangians have the form\footnote{The use of derivatives in
  Lagrangian does not change the main conclusions. We are interested in a fixed
  isospin, so isotopic indices are omitted.}:
\begin{alignat*}{2}
  \mathscr{L}_{int}&= \imath g_{1} \bar{N}_{1}(x) \gamma^{5} N(x) \phi (x)
  +\text{h.c.} &\quad&\text{for } J^{P}(N_1)={1/2}^{+},\\
  \mathscr{L}_{int}&= g_{2} \bar{N}_{2}(x) N(x) \phi (x) +\text{h.c.}
               &\quad&\text{for } J^{P}(N_2)={1/2}^{-} .
\end{alignat*}

In $n$-channel case, the scattering amplitude is a matrix of dimension $n$:
\begin{equation}\label{lab8}
  T = \bar{u}(p_{2},s_{2})  R  u(p_{1},s_{1}),
\end{equation}
where $\bar{u}(p_{2},s_2)$ and $u(p_{1},s_1)$ are four-component spinors,
corresponding to final and initial nucleon, and $R$ is matrix of the same
dimension $n$ consisting of the propagator and coupling constants.

In the two-channel approximation ($\pi N$ and $\eta N$ channel) matrix $R$ is of
the form:
\begin{equation}\label{rmatr}
  R= -
    \begin{pmatrix}
      \imath g_{1,\pi}\gamma^{5} & g_{2,\pi} \\
      \imath g_{1,\eta}\gamma^{5} & g_{2,\pi}
    \end{pmatrix}\times
  G\times
    \begin{pmatrix}
      \imath g_{1,\pi}\gamma^{5} & \imath g_{1,\eta}\gamma^{5}\\
      g_{2,\pi} & g_{2,\eta}
    \end{pmatrix},
\end{equation}
and generalization for $n$ channels and $m$ mixed states is obvious. Here $G$ is
dressed propagator \eqref{lab7} and we have introduced the short notations for
coupling constants: $g_{1,\pi}\equiv g_{N_{1} \pi N}$, $g_{2,\pi}\equiv
g_{N_{2}\pi N}$.\footnote{The matrix of coupling constants in the general case
  is a rectangular matrix. Note that the form of our amplitudes \eqref{lab8},
  \eqref{rmatr} similar to the multi-channel approach of Carnegie-Melon-Berkeley
  group \cite{Cut79}, the difference is in another form of the matrix propagator
  and vertex.}

After some algebra the matrix $R$ turns into into the standard form
\begin{equation}
  R= \Lambda^{+} R_1 + \Lambda^{-} R_2,
\end{equation}
where $R_1$ and $R_2$ are dimension 2 matrices. Note that the $\gamma^5$ matrix
has been disappeared after multiplication in \eqref{rmatr}, since parity is not
violated. After it we obtain from \eqref{lab8} the two-channel $s$- and $p$-
partial waves.

$s$-waves amplitudes (produced resonances have $J^{P}={1/2}^{-}$) in standard
notations have the form:
\begin{equation}\label{lab13}
  \begin{split}
    f_{s,+}(\pi N \rightarrow \pi N) &=
      \frac{(E_{1}+m_N)}{8\pi W \Delta_{2}} \Big[ g^{2}_{1,\pi}(W-m_{2}-\Sigma^{1}_{22})
      -g^{2}_{2,\pi}(-W-m_{1}-\Sigma^2_{11}) -\\
      &- \imath g_{1,\pi}g_{2,\pi}(\Sigma^{3}_{21}+\Sigma^{4}_{12}) \Big],\\
    f_{s,+}(\pi N \rightarrow \eta N) &=
      \frac{\sqrt{(E_{1}+m_N)(E_{2}+m_N)}}{8\pi W \Delta_{2}} \Big[ g_{1,\pi}g_{1,\eta}
      (W-m_{2}- \Sigma^{1}_{22}) -\\
      &- g_{2,\pi}g_{2,\eta}(-W-m_{1}-\Sigma^{2}_{11})
       -\imath g_{2,\eta} g_{1,\pi}\Sigma^{3}_{21}- \imath g_{1,\eta} g_{2,\pi}\Sigma^{4}_{12}\Big] ,\\
    f_{s,+}(\eta N \rightarrow \eta N) &=
      \frac{(E_{2}+m_N)}{8\pi W \Delta_{2}} \Big[ g^{2}_{1,\eta}(W-m_{2}-\Sigma^{1}_{22})
      -g^{2}_{2,\eta}(-W-m_{1}-\Sigma^{2}_{11})-\\
      &- \imath g_{1,\eta}g_{2,\eta}(\Sigma^{3}_{21}+\Sigma^{4}_{12}) \Big],\\
    \Delta_{2}&=\big(-W-m_1-\Sigma^{2}_{11}\big)
      \big(W-m_2-\Sigma^{1}_{22}\big)-\Sigma^{4}_{12}\Sigma^{3}_{21},
  \end{split}
\end{equation}
where $E_{1}$ and $E_{2}$ are nucleon energy in the c.m.s. for $\pi N$ and $\eta
N$ respectively.

For comparison, we write down the amplitude $\pi N \rightarrow \pi N$ in a tree
approximation:
\begin{equation}\label{lab15}
  f_{s,+}^{\text{tree}}(\pi N \rightarrow \pi N) =
    \dfrac{(E_{1}+m_N)}{8 \pi W }
  \Big[\dfrac{g^{2}_{1,\pi}}{(-W-m_{1})}-\dfrac{g^{2}_{2,\pi}}{(W-m_{2})}\Big].
\end{equation}

Simultaneous calculation of $p$-wave amplitudes
($J^{P}={1/2}^{+}$) gives:
\begin{equation}\label{lab14}
  \begin{split}
    f_{p,-}(\pi N \rightarrow \pi N) &=
       -\frac{(E_{1}-m_N)}{8\pi W \Delta_{1}} \Big[ g^{2}_{1,\pi}(-W-m_{2}-\Sigma^{2}_{22})
       -g^{2}_{2,\pi}(W-m_{1}-\Sigma^{1}_{11})-\\
      &- \imath g_{1,\pi}g_{2,\pi}(\Sigma^{4}_{21}+\Sigma^{3}_{12}) \Big],\\
    f_{p,-}(\pi N \rightarrow \eta N) &=
       -\frac{\sqrt{(E_{1}-m_N)(E_{2}-m_N)}}{8\pi W \Delta_{1}} \Big[ g_{1,\pi}g_{1,\eta}(-W-m_{2}-
        \Sigma^{2}_{22}) -\\
       &- g_{2,\pi}g_{2,\eta}(W-m_{1}-\Sigma^{1}_{11}) - \imath g_{2,\eta} g_{1,\pi}\Sigma^{4}_{21}
        -\imath g_{1,\eta} g_{2,\pi}\Sigma^{3}_{12} \Big], \\
    f_{p,-}(\eta N \rightarrow \eta N) &=
      -\frac{(E_{2}-m_N)}{8\pi W \Delta_{1}} \Big[ g^{2}_{1,\eta}(-W-m_{2}W-\Sigma^{2}_{22})
     -g^{2}_{2,\eta}(W-m_{1}-\Sigma^{1}_{11})- \\
    & - \imath g_{1,\eta}g_{2,\eta}(\Sigma^{4}_{21}+\Sigma^{3}_{12}) \Big],\\
    \Delta_{1}&=\big(W-m_1-\Sigma^{1}_{11}\big)\big(-W+m_2-\Sigma^{2}_{22}\big)-\Sigma^{3}_{12}\Sigma^{4}_{21}.
  \end{split}
\end{equation}

In tree approximation:
\begin{equation}\label{lab16}
  f_{p,-}^{\text{tree}}(\pi N \rightarrow \pi N) =
    \dfrac{(E_{1}-m)}{8\pi W } \Big[ -\dfrac{g^{2}_{1,\pi}}{(W-m_{1})}+
    \dfrac{ g^{2}_{2,\pi}}{(-W-m_{2})}\Big].
\end{equation}

One could convince oneself that the constructed partial amplitudes satisfy the
multi-channel unitary condition:
\begin{equation}\label{unit}
  \Im f_{ij}= \sum_{k}\ \abs{\mathbf{p}_k}\ f_{ik}\cdot f_{kj}^*,
\end{equation}
where $\mathbf{p}_k$ is the c.m.s. spatial momentum of particles in $k$-th
intermediate states.

The self-energy (before renormalization) is expressed through the components of
the standard loop functions $\Sigma_{\pi}(W)$ and $\Sigma_{\eta}(W)$:
\begin{equation}\label{Sigma}
  \begin{split}
    \Sigma^{1}_{11} &=-g^{2}_{1,\pi}\Sigma^{2}_{\pi}-g^{2}_{1,\eta}\Sigma^{2}_{\eta},\\
    \Sigma^{2}_{11} &=-g^{2}_{1,\pi}\Sigma^{1}_{\pi}-g^{2}_{1,\eta}\Sigma^{1}_{\eta},\\
    \Sigma^{1}_{22} &=g^{2}_{2,\pi}\Sigma^{1}_{\pi}+g^{2}_{2,\eta}\Sigma^{1}_{\eta},\\
    \Sigma^{2}_{22} &=g^{2}_{2,\pi}\Sigma^{2}_{\eta}+g^{2}_{2,\pi}\Sigma^{2}_{\eta},\\
    \Sigma^{3}_{12} &=\imath g_{1,\pi}g_{2,\pi}\Sigma^{2}_{\pi}+\imath g_{1,\eta}g_{2,\eta}\Sigma^{2}_{\eta},\\
    \Sigma^{4}_{12} &=\imath g_{1,\pi}g_{2,\pi}\Sigma^{1}_{\pi}+\imath g_{1,\eta}g_{2,\eta}\Sigma^{1}_{\eta},\\
    \Sigma^{3}_{21} &=\Sigma^{4}_{12},\\
    \Sigma^{4}_{21} &=\Sigma^{3}_{12},\\
  \end{split}
\end{equation}
where function $\Sigma_{\pi}(p)$ corresponding, for example, $\pi N$
intermediate state has the form:
\begin{equation*}
  \Sigma_{\pi}(p)=\dfrac{\imath}{(2\pi)^{4}} \int {\dfrac{d^{4}k}{(
      \widehat{p}-\widehat{k}-m_N )( k^{2}-m_{\pi}^{2})}} = A(p^2)+\hat{p}B(p^2)
  = \Lambda^{+}\Sigma^{1}_{\pi}(W)+\Lambda^{-}\Sigma^{2}_{\pi}(W).
\end{equation*}

It is convenient to calculate first $A$ and $B$ and then pass to the projections
$\Sigma^{1,2}$. So, we calculate discontinuities using Landau--Cutkosky rule:
\begin{equation*}
  \begin{split}
    \Delta A(p^2) &=-\imath\frac{m_{N}\abs{\mathbf{p}_\pi}}{4\pi W}, \\
    \Delta B(p^2) &=-\imath\frac{\abs{\mathbf{p}_\pi}(p^{2}+m^{2}_{N}-m^{2}_{\pi})}{8\pi p^2 W },
  \end{split}
\end{equation*}
then restore functions $A(p^2)$ and $B(p^2)$ through dispersion relation, and
finally calculate $\Sigma^{1,2}$:
\begin{equation*}
  \begin{split}
    \Sigma^{1} &=A(W^{2})+W B(W^{2}),\\
    \Sigma^{2} &=A(W^{2})-W B(W^{2}).
  \end{split}
\end{equation*}

Let us write down the imaginary parts of $\Sigma^{1,2}$:
\begin{equation}\label{impr}
  \begin{split}
    \Im\Sigma^1_\pi &= -\frac{\abs{\mathbf{p}_\pi}(E_1+m_N)}{8\pi W},\\
    \Im\Sigma^2_\pi &= \phantom{-}\frac{\abs{\mathbf{p}_\pi}(E_1-m_N)}{8\pi W},
  \end{split}
\end{equation}
where $\mathbf{p}_\pi$ is momentum of pion in the c.m.s.

Recall that decomposition coefficients in the projection basis are related with
each other by the substitution $W \rightarrow -W$. So, to renormalize the
self-energy, it is sufficient to define an exact form of $\Sigma^{1}(W)$ and
$\Sigma^{3}(W)$, then the components $\Sigma^{2}(W)$, $\Sigma^{4}(W)$ are fixed
by symmetry.  We will use the on-mass-subtraction method of renormalization of
resonance contribution \cite{Aok82,Den93}.

Subtraction conditions for the self-energy included in the $s$-wave amplitudes
have the form\footnote{Note that the non-diagonal self-energy terms have
  additional factor $i$ -- see \eqref{Sigma}.}:
\begin{equation}\label{Subs}
  \begin{split}
  \Re\Sigma^{1}_{22}(W)\quad &\text{has zero of second order at $W=m_{2}$},\\
  \Re\Sigma^{2}_{11}(W)\quad &\text{has zero of second order at $W=-m_{1}$},\\
  \Im\Sigma^{3}_{21}(W)\quad &\text{has zeros at $W=-m_{1}$ and $W=m_{2}$} .\\
  \end{split}
\end{equation}

After it the $p$-wave amplitudes are determined by replacing $W \rightarrow -W$,
as it was mention above.\footnote{The known McDowell's symmetry \cite{Mcd},
  connecting different partial waves $f_{l,+}(-W)=-f_{l+1,-}(W)$, is a
  consequence of the symmetry properties of coefficients in the projection
  basis:\ $G^2(W)=G^1(-W)$, $G^4(W)=G^3(-W)$.}

Recall also the relationships between coupling constants and decay widths in the
absence of mixing:
\begin{equation}\label{sigma}
  \begin{split}
    \Gamma\big(N_1(1/2^{+})\rightarrow \pi N\big) &=\dfrac{g^{2}_{1,\pi}}{4\pi}\cdot
    \dfrac{\abs{\mathbf{p}_\pi}(E_{1}-m_{N})}{ M}, \\
    \Gamma\big(N_2(1/2^{-})\rightarrow \pi N\big) &=\dfrac{g^{2}_{2,\pi}}{4\pi}\cdot
    \dfrac{\abs{\mathbf{p}_\pi}(E_{1}+m_{N})}{ M} .
  \end{split}
\end{equation}

\subsection{Comparison with the $K$-matrix}

The usual definition of the $K$-matrix is:
\begin{equation}\label{labK1}
  T=K\big(I-\imath pK\big)^{-1},
\end{equation}
where $T$ is matrix of partial amplitudes, $p$ is diagonal matrix consisting of
c.m.s. momenta:
\begin{equation}
  p =
    \begin{pmatrix}
      \abs{\mathbf{p}_\pi}, & 0\\
      0, & \abs{\mathbf{p}_\eta}
    \end{pmatrix}.
\end{equation}

$K$-matrix representation by construction satisfies the unitary
condition. Usually, the $K$-matrix represents a set of poles and,
possibly, some smooth contributions.

Another variant is the analytical $K$-matrix (for example,
\cite{Bab76,Arn85})
\begin{equation}\label{labK2}
  T=K(I-CK)^{-1}.
\end{equation}

The presentation \eqref{labK2} differs from the standard $K$-matrix
\eqref{labK1} by the presence of a matrix $C$ consisting of loops, whose
imaginary part is equal to the matrix $p$.

It is convenient to rewrite \eqref{labK2} in terms of inverse matrix:
\begin{equation}\label{K3}
  T^{-1}=K^{-1}-C.
\end{equation}

It turns out that the our partial amplitudes \eqref{lab13}, \eqref{lab14} can be
represented in the form \eqref{labK2}, \eqref{K3}. As an example consider the
two-channel $s$-wave amplitudes $f_{s,+}$ \eqref{lab13} and use the self-energy
in form of \eqref{Sigma}, without subtraction polynomials. Calculating the
inverse matrix of the amplitudes we find that, in accordance with \eqref{K3}, it
consists of a loop matrix and pole matrix
\begin{equation}
  T^{-1}=K^{-1}+\dfrac{1}{8 \pi W}
    \begin{pmatrix}
      \dfrac{\Sigma_{11}}{E_{1}+m}, & 0 \\
      0, & \dfrac{\Sigma_{22}}{E_{2}+m}
    \end{pmatrix}.
\end{equation}

Our amplitudes \eqref{lab13}, \eqref{lab14} lead to a pole contributions of the
form:
\begin{multline}\label{ourk}
  K = \frac{-1}{8 \pi W}
    \begin{pmatrix}
      \sqrt{E_{1}+m}, & 0 \\
      0, & \sqrt{E_{2}+m}
    \end{pmatrix} \times
    \begin{pmatrix}
      \dfrac{g^{2}_{1,\pi}}{W+m_{1}}+\dfrac{g^{2}_{2,\pi}}{W-m_{2}},
      & \dfrac{g_{2,\pi} g_{2,\eta}}{W-m_{2}}+\dfrac{g_{1,\pi} g_{1,\eta}}{W+m_{1}} \\
      \dfrac{g_{1,\pi} g_{1,\eta}}{W+m_{1}}+\dfrac{g_{2,\pi}
        g_{2,\eta}}{W-m_{2}}, &
      \dfrac{g^{2}_{1,\eta}}{W+m_{1}}+\dfrac{g^{2}_{2,\eta}}{W-m_{2}}
    \end{pmatrix} \times \\ \times
    \begin{pmatrix}
      \sqrt{E_{1}+m}, & 0 \\
      0, & \sqrt{E_{2}+m}
    \end{pmatrix}.
\end{multline}

In resonance phenomenology $K$-matrix contains a set of poles, corresponding to
bare states. The main feature of our $K$-matrix \eqref{ourk} is the presence of
poles both with positive and negative energy. If the self-energy in addition to
\eqref{Sigma} contains the subtraction polynomials, it leads only to
redefinition of the poles positions in $K$-matrix (i.e. $K$-matrix masses
$m_{1},m_{2}$).

We can see that our multi-channel amplitudes \eqref{lab13}, \eqref{lab14} can be
reduced to some specific version of the analytical $K$-matrix parametrization.

\subsection{Estimates of observed effects}

Let us use our amplitudes \eqref{lab13},\eqref{lab14} to calculate $\pi N$
partial $s$- and $p$-waves, where baryons $J^P = 1/2^{\pm}$ can be produced. We
are interested here only in estimates of the observed effects, so we restrict
ourselves by the single-channel approach and fix the parameters (masses and
coupling constants) from rough correspondence to parameters of the observed
baryon resonances $I=1/2$
\begin{equation}\label{lab20}
  \begin{aligned}
     P_{11}(1440),\quad J^P = 1/2^+ :\; M_1=1.440\,\text{GeV}, \\
      \Gamma_1=300\,\text{MeV} \Rightarrow g_{1,\pi}=13.0\,\text{GeV} \\
     S_{11}(1535),\quad J^P = 1/2^- 7:\; M_2=1.535\,\text{GeV}, \\
      \Gamma_2=150\,\text{MeV} \Rightarrow g_{2,\pi}=1.77\,\text{GeV}.
  \end{aligned}
\end{equation}

For estimates we used the relations \eqref{sigma} of the widths and coupling
constants in the absence of mixing \eqref{sigma}.

\begin{figure}[h]
  \begin{center}
    \includegraphics[width=8cm]{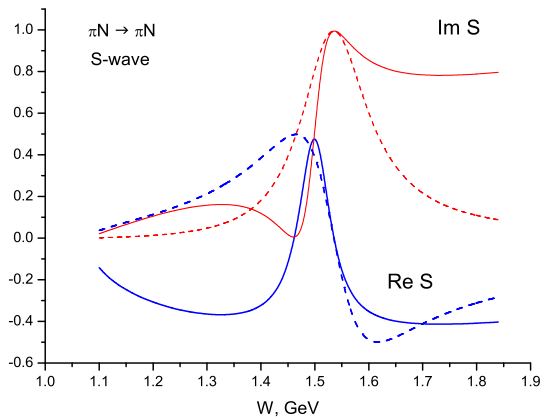}
  \end{center}
  \caption{\label{estS} The results of calculations of $\pi N $ $s$-wave partial
    wave. Solid lines correspond to the real and imaginary parts of our partial
    amplitude \eqref{lab13}, \eqref{lab14} in the single-channel approach with
    the parameters \eqref{lab20}. Dashed lines correspond to our amplitudes,
    neglecting the mixing effect: $\Sigma_{12}=\Sigma_{21}=0$.  All variants of
    amplitudes satisfy the single-channel unitary condition $\Im S= \abs{S}^2$.}
\end{figure}

\begin{figure}[h]
  \begin{center}
    \includegraphics[width=8cm]{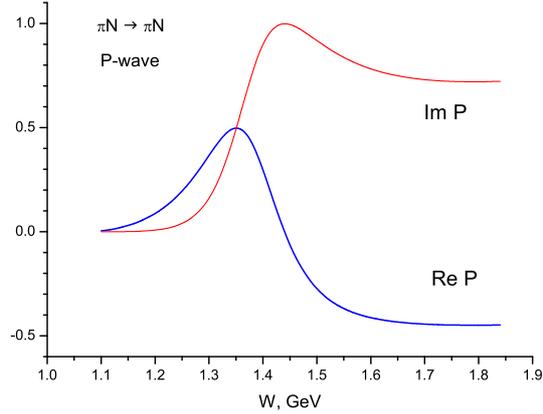}
  \end{center}
  \caption{\label{estP} The real and imaginary parts of $\pi N $ partial
    $p$-wave.  Notations are the same as in Fig.~\ref{estS}. For $p$-wave the
    solid and dashed lines coincide with each other.}
\end{figure}

The results of calculations of $\pi N$ partial waves are shown at
Figs.~\ref{estS}, \ref{estP}.

It turns out that the discussed OPF-mixing leads to noticeable effects only in
$s$-wave, while its influence in $p$-wave is much less and does not seen at
graphics. This feature is explained by the values of the coupling constants in
\eqref{lab20} $\abs{g_{2,\pi}}\ll\abs{g_{1,\pi}}$ and may be seen at qualitative
level from the tree amplitudes \eqref{lab15}, \eqref{lab16}. Since we have
normalized the coupling constants on the resonance width, inequality between the
coupling constants is a consequence of the inequality between the $s$- and
$p$-wave phase volumes.

We see that the discussed mixing effect generate the (unitary) interference
picture ``resonance + background'' in the $s$-wave.  In this case the $s$-wave
background contribution originates from the $p$-wave resonance and gives the
negative contribution to $s$-wave phase shift. This fact can be seen from
Fig.~\ref{estS} and from eq. \eqref{lab13}.

\begin{figure}[ht]
  \begin{center}
    \includegraphics[width=13cm]{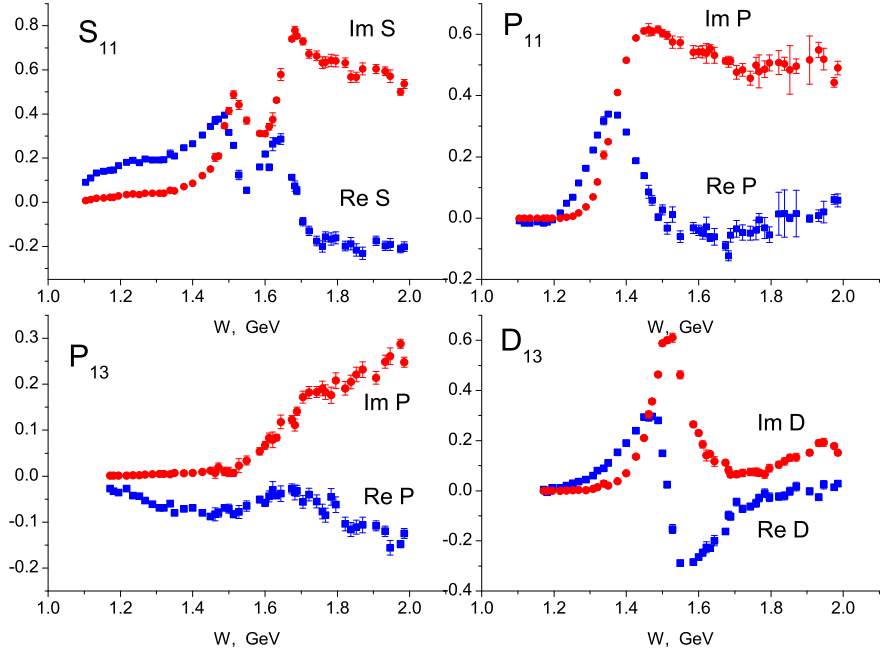}
  \end{center}
  \caption{\label{GWU} The results of partial wave analysis \cite{Arn06} for
    $\pi N $ scattering amplitudes with isospin $I=1/2$ (current solution).
    Partial waves satisfy the unitary condition $\Im T=
    \abs{T}^2+(1-\eta^2)/4$.}
\end{figure}

Fig.~\ref{GWU} demonstrates the results of partial wave analysis \cite{Arn06}
for lowest $\pi N$ amplitudes with isospin $I=1/2$.  The discussed effect leads
to hard correlation between pair of partial waves. From physical point of view
the most interesting is the pair of waves $S_{11}, P_{11}$; recall that in the
$J^P=1/2^+$ sector there exist up to now the problems of physical interpretation
of the observed states and their correspondence with quark models, see
e.g. discussions in \cite{Cap00,Kre00,Sar08,Ces08}. But this pair of partial
waves is not the simplest place for identification of the discussed OPF-mixing
effect. The reasons are the old problem with Roper resonance (non-standard form
of $1/2^+$ state) and the existence of several states in $1/2^-$ channel.

But if to look at the partial waves $P_{13},D_{13}$, where resonances
$3/2^{\pm}$ are produced, here we observe the more evident situation, which is
qualitatively consistent with our expectations, shown at Figs.~\ref{estS},
\ref{estP}. Namely: in the $d$-wave we see a single resonance, whereas in the
$p$-wave there is a visible interference of resonance with a background.
Moreover, in accordance with our expectations for interference picture, the
background in the $p$-wave is evidently negative -- see Fig.~\ref{GWU}. So this
pair of partial waves $P_{13},D_{13}$ looks as a most suitable place for
identification of the discussed mixing effect.

\section{OPF-mixing for baryons $3/2^{\pm}$}
\label{sec:mixing}
The above discussion was devoted to mixing of two Dirac fields of opposite
parities, the same effect arises for vector-spinor fields $\Psi^\mu$, which
describe the spin-3/2 particles. We want to obtain the hadron partial
amplitudes, which take into account the discussed effect, and to use them for
description of results of $\pi N$ partial wave analysis.

The details of calculations of the spin-3/2 baryons production are given in the
appendix~\ref{ap2}. Here we present only the results of calculations: the hadron
partial amplitudes in two-channel ($\pi N$, $\eta N$) approach (compare them
with spin-1/2 case \eqref{lab13}, \eqref{lab14}).

$p$-wave amplitudes ($J^{P}={3/2}^{+}$) have the form:
\begin{equation}\label{lab21}
  \begin{split}
    f_{p,+}(\pi N \rightarrow \pi N) &=
      \abs{\mathbf{p}_\pi}^{2}\frac{(E_{1}+m)}{24\pi W \Delta_{2}}
      \Big[ g^{2}_{1,\pi}(W-m_{2}-\Sigma^{1}_{22})
      - g^{2}_{2,\pi}(-W-m_{1}-\Sigma^2_{11})+ \\
     &+ \imath g_{1,\pi}g_{2,\pi}(\Sigma^{3}_{21}+\Sigma^{4}_{12}) \Big],\\
    f_{p,+}(\pi N \rightarrow \eta N) &=
      \abs{\mathbf{p}_\pi}\abs{\mathbf{p}_\eta}
      \frac{\sqrt{(E_{1}+m)(E_{2}+m)}}{24\pi W \Delta_{2}}
      \Big[g_{1,\pi}g_{1,\eta}(W-m_{2}- \Sigma^{1}_{22})-\\
     & -g_{2,\pi}g_{2,\eta}(-W-m_{1}-\Sigma^{2}_{11})
       +\imath g_{1,\pi}g_{2,\eta}\Sigma^{4}_{12} + \imath g_{2,\pi}g_{1,\eta}\Sigma^{3}_{21} \Big],\\
    f_{p,+}(\eta N \rightarrow \eta N) &=
      \abs{\mathbf{p}_\eta}^{2}\frac{(E_{2}+m)}{24\pi W \Delta_{2}}
      \Big[g^{2}_{1,\eta}(W-m_{2}-\Sigma^{1}_{22})
      -g^{2}_{2,\eta}(-W-m_{1}-\Sigma^{2}_{11})+ \\
    & + \imath g_{1,\eta}g_{2,\eta}(\Sigma^{3}_{21}+\Sigma^{4}_{12}) \Big],\\
    \Delta_{2}&=\big(-W-m_{1}-\Sigma^2_{11}\big)\big(W-m_{2}-\Sigma^{1}_{22}\big)- \Sigma^{4}_{12}\Sigma^{3}_{21}.
  \end{split}
\end{equation}

$d$-wave amplitudes ($J^{P}={3/2}^{-}$):
\begin{equation}\label{lab22}
  \begin{split}
    f_{d,-}(\pi N \rightarrow \pi N) & =
      \abs{\mathbf{p}_\pi}^{2}\frac{(E_{1}-m)}{24\pi W \Delta_{1}} \Big[
      -g^{2}_{1,\pi}(-W-m_{2}-\Sigma^{2}_{22})
      + g^{2}_{2,\pi}(W-m_{1}-\Sigma^{1}_{11}) - \\
     &- \imath g_{1,\pi}g_{2,\pi}(\Sigma^{4}_{21}+\Sigma^{3}_{12}) \Big],\\
    f_{d,-}(\pi N \rightarrow \eta N) & =
      \abs{\mathbf{p}_\pi}\abs{\mathbf{p}_\eta}
      \frac{\sqrt{(E_{1}-m)(E_{2}-m)}}{24\pi W \Delta_{1}}
      \Big[ -g_{1,\pi}g_{1,\eta}(-W-m_{2} - \Sigma^{2}_{22}) + \\
     &+ g_{2,\pi} g_{2,\eta}(W-m_{1}-\Sigma^{1}_{11})
      - \imath g_{1,\pi} g_{2,\eta}\Sigma^{3}_{12} - \imath g_{2,\pi} g_{1,\eta}\Sigma^{4}_{21} \Big],\\
    f_{d,-}(\eta N \rightarrow \eta N) & =
      \abs{\mathbf{p}_\eta}^{2}\frac{(E_{2}-m)}{24\pi W \Delta_{1}}
      \Big[-g^{2}_{1,\eta}(-W-m_{2}-\Sigma^{2}_{22})
      + g^{2}_{2,\eta}(W-m_{1}-\Sigma^{1}_{11}) -\\
     &- \imath g_{1,\eta}g_{2,\eta}(\Sigma^{4}_{21}+\Sigma^{3}_{12}) \Big],\\
    \Delta_{1}&=\big(W-m_{1}-\Sigma^1_{11}\big)\big(-W-m_{2}-\Sigma^{2}_{22}\big)-\Sigma^{3}_{12}\Sigma^{4}_{21}.
  \end{split}
\end{equation}
where $E_{1}$ and $E_{2}$ are nucleon energies for $\pi N$ and $\eta N$ states
respectively.

The obtained $p$ and $d$ partial amplitudes satisfy the two-channel unitary
condition \eqref{unit}.

Besides, we should take into account the $W$-dependent form-factor in a vertex
(the so called centrifugal barrier factor). There is no common opinion in
literature concerning its form, we take it in two-parameter form:
\begin{equation}\label{FF}
  g \rightarrow g \cdot F(W^2) = g \cdot \dfrac{1+a M^{2}+b M^4}{1+a W^{2}+b W^4}.
\end{equation}

The partial amplitudes \eqref{lab21}, \eqref{lab22}, which take into account the
OPF-mixing, are written in two-channel approach.  But in fact in considered
region of energy $W< 2$ GeV there exist at least five open channels, the most
essential are the $(\pi\pi)_S N$ and $\pi \Delta$ channels. In this situation we
follow the way suggested in \cite{Bat95,Cec08,Ces08}: we restrict ourselves by
the three-channel approach ($\pi N$, $\eta N$ and $\sigma N$). As for third
channel ($\sigma N= \pi\pi N$), it is considered as some ``effective'' channel and
its threshold may be a free parameter in a fit.

Three-channel amplitudes may be obtained from the formulae \eqref{R32},
\eqref{V32} in appendix~\ref{ap2}, but they are rather cumbersome so we did not
write down them.  For our local purpose of the description of $\pi N \to \pi N$
amplitudes, it is sufficient to use formulae \eqref{lab21}, \eqref{lab22}. The
only difference will appear in the self-energy, where we should add the third
channel in the similar manner. We use the same procedure of loop renormalization
as for spin $1/2$, see \eqref{Subs}.

First of all let's try to describe the ${P_{13}}$, ${D_{13}}$ separately. We
found that, in accordance with our estimates for spin-1/2 case, the OPF-mixing
is more essential for lowest $l$ wave ${P_{13}}$.

\begin{figure}[h]
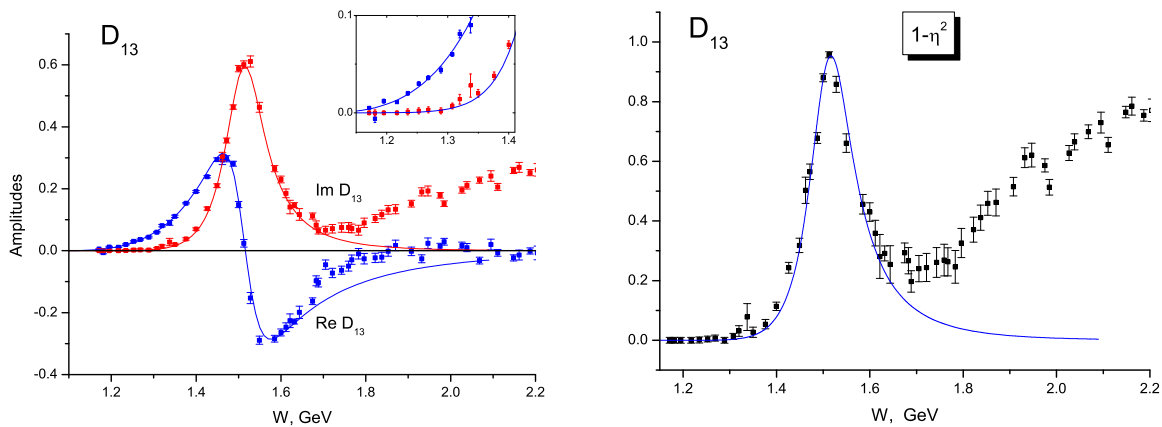

  \includegraphics[width=8cm]{\imagepath{D13_single}}
  \includegraphics[width=8cm]{\imagepath{D13_inel}}
  \caption{\label{D13_s}Left: $D_{13}$ partial wave of $\pi N$ scattering
    \cite{Arn06} and results of fit by our formulae with $\pi N$ and $\sigma N$
    channels ($W < 1.7$ GeV). Right: inelasticity from PWA \cite{Arn06} and our
    curve, corresponding to left panel.}
\end{figure}

Results of $D_{13}$ fitting by formulae \eqref{lab22} in two-channel ($\pi N$,
$\sigma N$) approach are shown at Fig.~\ref{D13_s}. We restricted the energy
interval by $W<1.7$ GeV since at higher energy there appears some
additional smooth contribution --- it is seen well from $1-\eta^2$ behaviour.
As for mass of ``effective'' $\sigma$-meson, fit leads to rather low value
$m_\sigma \leq 0.3$ GeV. From other side, the d-wave threshold generates
rather smooth contribution in amplitude and is defined badly from data. So we
fix it by $m_\sigma = 280$ MeV in the following.

Fit of real and imaginary parts of $D_{13}$ gives:
\begin{multline}\label{fitD}
  m_1=1.5161\pm 0.0005\,\text{GeV},\quad g_{1,\pi}=20.23\pm 0.10\,\text{GeV},
  \quad g_{1,\sigma}=21.60\pm 0.25\,\text{GeV},\\
  \chi^{2}/\text{DOF} = 213/59 .
\end{multline}
Parameters of form-factor from $D_{13}$ wave:
\begin{equation}\label{FF1}
  a=-1.005\pm 0.009\,\text{GeV}^{-2},\quad b=0.434\pm 0.021\,\text{GeV}^{-4}
\end{equation}

\begin{figure}[h]
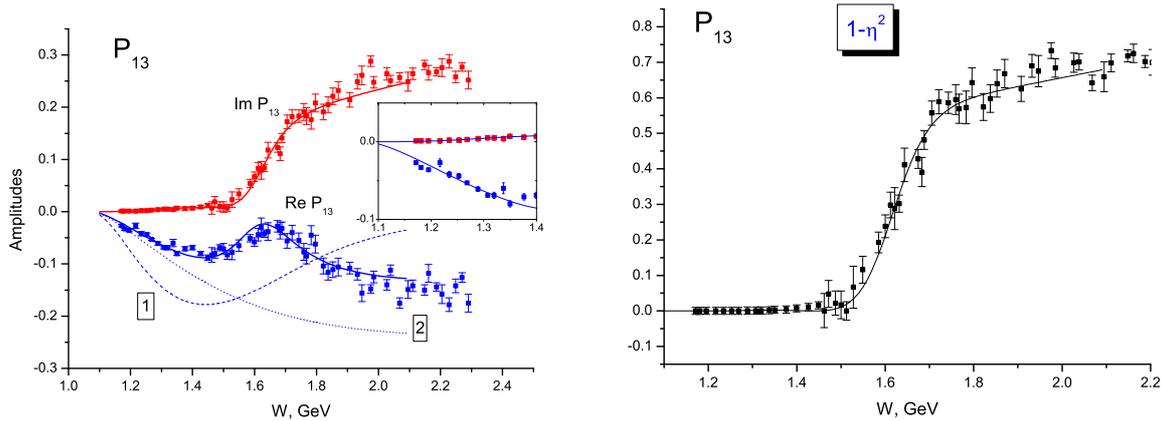

  \includegraphics[width=8cm]{\imagepath{P13_single}}
  \includegraphics[width=8cm]{\imagepath{P13_inel}}
  \caption{\label{P13_single} $P_{13}$ partial wave of $\pi N$ scattering
    \cite{Arn06} and results of fit by our formulae with $\pi N$ and $\sigma N$
    channels ($W < 2.0$ GeV). Parameters of $D_{13}$ resonance are fixed by
    \eqref{fitD}.  Curves 1 and 2 show the real part of background contribution
    from $D_{13}$ resonance ($g_{2,\pi}=g_{2,\sigma}=0$) with form-factors
    \eqref{FF1} and \eqref{FF2}.  Right: inelasticity from PWA \cite{Arn06} and
    our curve, corresponding to left panel.}
\end{figure}

Now we can describe $P_{13}$ at fixed parameters \eqref{fitD} of $D_{13}$
resonance. Results are shown at Fig.~\ref{P13_single}.

\begin{multline}\label{fitP}
  m_2=1.721\pm 0.005\,\text{GeV},\quad g_{2,\pi}=3.73\pm 0.10\,\text{GeV},
  \quad g_{2,\sigma}=9.23\pm 0.25\,\text{GeV},\\
  \chi^{2}/\text{DOF} = 210/91.
\end{multline}
Parameters of form-factor from $P_{13}$ wave:
\begin{equation}\label{FF2}
  a=1.51\pm 0.30\,\text{GeV}^{-2},\quad b=0.001\pm 0.017\,\text{GeV}^{-4}
\end{equation}

We observe that both fits are consistent with each other in parameters of
resonances, except for the vertex form-factor. The
obtained parameters do not contradict to values of masses and branching ratios
of $D_{13}(1520)$, $P_{13}(1720)$ in RPP tables \cite{RPP}.

As for $\eta N$ channel: PWA results for $P_{13}$ wave does not require this
coupling. For $D_{13}$ situation is unstable: inclusion of this coupling leads
to unphysical big coupling constants. But close inspection shows that this is
effect of another threshold with higher mass. So we will restrict ourselves by
the two-channel approach.

\begin{figure}[h]
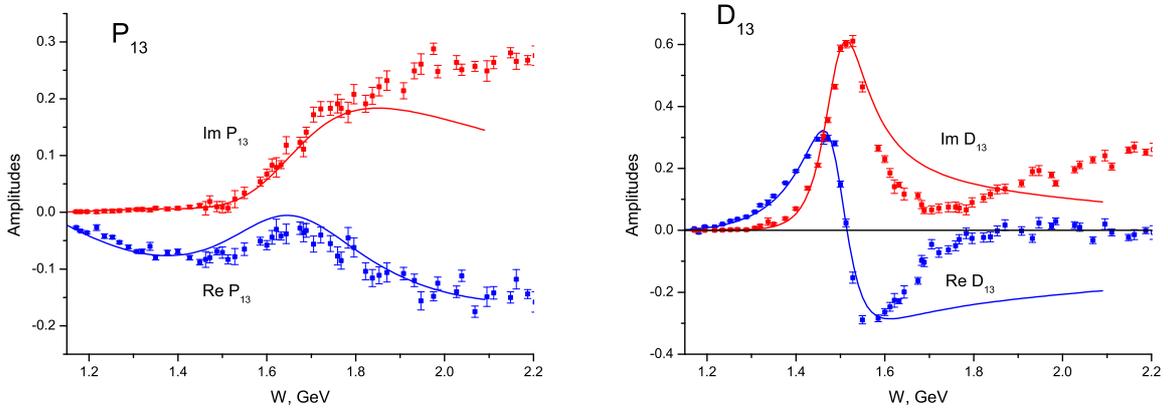

  \includegraphics[width=8cm]{\imagepath{P13_joint}}
  \includegraphics[width=8cm]{\imagepath{D13_joint}}
  \caption{\label{PDjoint}Example of joint description of $P_{13}$ (W < 2.0 GeV) and $D_{13}$
    ($W < 1.6$ GeV) partial waves by our formulae with OPF-mixing in two-channel
    approach. In this case $\chi^{2}/\text{DOF} = 1220/139$. }
\end{figure}

Figs.~\ref{D13_s}, \ref{P13_single} demonstrate that fit of $D_{13}$ and $D_{13}$
separately leads to rather good quality of description.  As for joint fit -- it
gives only qualitative description, as it seen from Fig.\ref{PDjoint}. For
better quality it needs ``fine tuning'', first of all it should include:
\begin{itemize}
\item More accurate description of $(\pi\pi)N$ channel;
\item Account of smooth contribution in $D_{13}$ wave -- see Fig.~\ref{D13_s};
\item Better understanding of role and properties of the vertex form-factor.
  The observed disagreement may be related with above items.
\end{itemize}

Thus we can see that the considered mixing of the opposite parities fermion
fields leads to the sizeable effects for baryon production and may be identified
in production of baryon resonances $3/2^{\pm}$ in $\pi N$ scattering.

\section{Conclusion}
\label{sec:conclus}

In present paper we have analyzed the mixing effect, specific for fermions, when
two fermion fields of opposite parities are mixed at loop level. For fermions it
is possibly even if the parity is conserved in a vertex. As a result we have a
matrix propagator of unusual form \eqref{lab7}, which contains $\gamma^5$
contributions. But since parity is conserved in vertexes, the $\gamma^5$ matrix
disappears after multiplication by the vertexes, and we get the amplitudes
containing the resonance and background contributions. Note that as a result of
solving the Dyson--Schwinger equations we automatically obtain the unitary
amplitudes.

The derived amplitudes resemble in structure the analytical $K$-matrix. The most
significant difference is the presence of poles both of positive and negative
energies in our amplitudes.

If to say about resonance phenomenology, we have a pair of partial waves with
strongly correlated parameters, namely, the resonance in one partial wave is
connected with background contribution in another wave. The discussed effect is
most essential for partial wave with smaller orbital momentum $l$, thit is a
consequence of inequality of phase volumes for different $l$.

As for manifestation of this effect in $\pi N$ scattering, the most simple
physical example is connected with production of spin-$3/2$ resonances of
opposite parities and isospin $I=1/2$. We used the obtained amplitudes for
description of two $\pi N$ partial waves $\mathsf{P_{13}}$ and
$\mathsf{D_{13}}$. We can conclude that the discussed effect reproduces
naturally all the observed features of these partial waves but the joint
description of these partial waves needs fine tuning of their properties.

We suppose that the most interesting application of this effect is related with
the problem of Roper resonance $N(1440)$, $1/2^+$.  Recall that for these
quantum numbers there are still problems of physical interpretation of the
baryon states and their comparison with quark models. The effect of OPF-mixing
in this sector takes a more complicated form because of presence of several
states $1/2^-$ (see Fig.~\ref{GWU}) and non-standard form of the Roper resonance
$1/2^+$. But the above mentioned strong correlation between two partial waves
gives new possibilities for studying the properties of $N(1440)$.

\begin{center}
\textbf{Acknowledgements}
\end{center}
This work was supported in part by the program ``Development of Scientific
Potential in Higher Schools'' (project 2.2.1.1/1483, 2.1.1/1539) and by the
Russian Foundation for Basic Research (project No. 09-02-00749).

%%%%%%%%%%%%%%%%%%%%%%%%%%%%%%%%%%%%%%%%%%%
%%%%%%%%%%%%%%%% APPENDIX %%%%%%%%%%%%%%%%%
%%%%%%%%%%%%%%%%%%%%%%%%%%%%%%%%%%%%%%%%%%%
\newpage
\appendix

\section{Amplitudes of production of spin-$3/2$ resonances}
\label{ap2}

Let us write down the phenomenological Lagrangians of interaction of spin $3/2$
particles with $\pi N$ system.

For $J^P=3/2^+$ we have:
\begin{equation}
  \mathscr{L}=g_{R,\pi}\bar{\Psi}\vphantom{\Psi}_{\mu}(x) \Psi(x)\cdot\partial_{\nu}\phi(x) +
  \text{h.c.} .
\end{equation}

For $J^P=3/2^-$:
\begin{equation}
  \mathscr{L}=i g_{R,\pi}\bar{\Psi}\vphantom{\Psi}_{\mu}(x)
  \gamma^5 \Psi(x)\cdot\partial_{\nu}\phi(x) + \text{h.c.} .
\end{equation}
Here $\Psi_{\mu}$ is the vector-spinor Rarita-Schwinger field, isotopical
indices are omitted.

We are interested in the resonance contribution (the term of the leading spin
$s=3/2$ in this diagram).
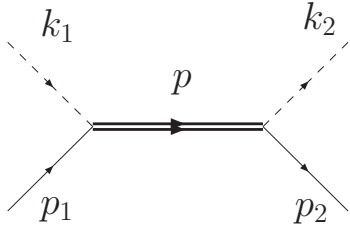
\begin{figure}[h]
  \begin{center}
    \includegraphics[width=5cm]{\imagepath{delta}}
  \end{center}
\end{figure}

Propagator of Rarita-Schwinger field has the form (see more in
\cite{Kal06,Kal04}):
\begin{equation}
  \begin{split}
    G^{\mu\nu}(p) &= \mathcal{P}^{\mu\nu}_1\cdot \bar{G}_1(W) +
    \mathcal{P}^{\mu\nu}_2 \cdot \bar{G}_2(W) + \text{($s=1/2$ contributions)},
  \end{split}
\end{equation}
where the basis elements are
\begin{equation}
  \mathcal{P}^{\mu\nu}_1=\Lambda^+ P^{\mu\nu}_{3/2},\quad
  \mathcal{P}^{\mu\nu}_2=\Lambda^- P^{\mu\nu}_{3/2}.
\end{equation}
The operator $P_{3/2}$ looks like \cite{Pvn}:
\begin{equation}\label{eq:oper}
    P_{3/2}^{\mu\nu} = g^{\mu\nu}-n_{1}^{\mu}n_{1}^{\nu}-n_{2}^{\mu}n_{2}^{\nu},
\end{equation}
where we have introduced the unit "vectors" orthogonal to each other:
\begin{equation}\label{eq:sv-uv}
  n_{1}^{\mu}=\frac{1}{\sqrt{3}p^{2}}(-p^{\mu}+\gamma^{\mu}\hat{p})\hat{p}, \quad
  n_{2}^{\mu}=\frac{p^{\mu}}{\sqrt{p^2}}, \quad
  (n_{i}\cdot n_{j})=\delta_{ij}.
\end{equation}

In the presence of parity violation or when considering the OPF-mixing the basis
in the sector $s=3/2$ must be supplemented by elements containing $\gamma^5$:
\begin{equation}
\begin{split}
  Q^{\mu\nu}_1 &=\mathcal{P}^{\mu\nu}_1, \quad\quad
  Q^{\mu\nu}_2 =\mathcal{P}^{\mu\nu}_2, \\
  Q^{\mu\nu}_3 &={\cal P}^{\mu\nu}_1\gamma^5 , \quad\quad
  Q^{\mu\nu}_4 ={\cal P}^{\mu\nu}_2 \gamma^5 .\\
  \end{split}
\end{equation}

Suppose we have two fields $\Psi^\mu$ of opposite parities. When taking into
account OPF-mixing the dressed propagator has the following decomposition:
\begin{equation}
  G^{\mu\nu}(p)=\sum_{M=1}^{4} Q^{\mu\nu}_M\cdot \bar{G}_M(W) + \text{($s=1/2$ contributions)},
\end{equation}
where $\bar{G}_M(W)$ being dimension 2 matrices are solutions of the matrix
Dyson--Schwinger equation.

Since the multiplicative properties of the operators $Q^{\mu\nu}_M$ are
completely consistent with the properties of the spin-$1/2$ operators (see
Table~\ref{multiplic}), the further calculations repeat $s=1/2$ ones. As a
result the matrix propagator looks similar to spin-$1/2$ case \eqref{lab7}.

Matrix amplitude has the form:
\begin{equation}
  T = \bar{u}(p_2,s_2) R u(p_1,s_1),
\end{equation}
where the matrix $R$ is constructed from the matrix of the propagator and vertex
matrices:
\begin{equation}\label{R32}
  R= -V^T\times
  \bigg( \sum_{M=1}^{4} k_2^\mu Q^{\mu\nu}_M k_1^\nu \cdot \bar{G}_M(W) \bigg)\times V .
\end{equation}
The vertex matrix in two-channel approximation looks like
\begin{equation}\label{V32}
  V = \begin{pmatrix}
      g_{1,\pi}\gamma^5 & g_{1,\eta}\gamma^5 \\
      ig_{2,\pi} & ig_{2,\eta}
    \end{pmatrix}.
\end{equation}
The self-energy
\begin{equation}\label{self32}
  \Sigma^{\mu\nu}=-V
  \begin{pmatrix}
    \hat{\Sigma}^{\mu\nu}_\pi & 0 \\
    0 & \hat{\Sigma}^{\mu\nu}_\eta
  \end{pmatrix} V^T + \text{subtraction},
\end{equation}
is expressed through the standard loop function corresponding to one of the
channels. For $\pi N$ channel this standard function has form:
\begin{equation}
    \hat{\Sigma}^{\mu\nu}_\pi = -i \int \dfrac{d^4 k}{(2\pi)^4}\dfrac{k^\mu
      k^\nu}{(\hat{p}-\hat{k}-m_N)(k^2-m_\pi^2)} = Q^{\mu\nu}_1\cdot
    \hat{\Sigma}^1_\pi + Q^{\mu\nu}_2\cdot \hat{\Sigma}^2_\pi + \text{($s=1/2$
      contributions)},
\end{equation}
and similarly for $\eta N$ the channel. An alternative decomposition of the loop
is
\begin{equation}
  \hat{\Sigma}^{\mu\nu}_\pi = (A_{\pi}(p^2)+\hat{p}B_{\pi}(p^2)) P_{3/2}^{\mu\nu} + \text{($s=1/2$ contributions)},
\end{equation}
so that
\begin{equation}
  \begin{split}
    \hat{\Sigma}^1_\pi(W) &=A_{\pi}(W^2)+W B_{\pi}(W^2),\\
    \hat{\Sigma}^2_\pi(W) &=A_{\pi}(W^2)-W B_{\pi}(W^2).
  \end{split}
\end{equation}

Imaginary parts are
\begin{equation}
  \begin{split}
    \Im A_{\pi} &=-\dfrac{\abs{\mathbf{p}_\pi}^3 m_N}{24\pi W}, \\
    \Im B_{\pi} &=-\dfrac{\abs{\mathbf{p}_\pi}^3(W^2+m_N^2-m_\pi^2)}{48\pi W^3},
  \end{split}
\end{equation}
and hence
\begin{equation}
  \begin{split}
    \Im\hat{\Sigma}^1_\pi &= -\dfrac{\abs{\mathbf{p}_\pi}^3(E_1+m_N)}{24\pi W},\\
    \Im\hat{\Sigma}^2_\pi &= \dfrac{\abs{\mathbf{p}_\pi}^3 (E_1-m_N)}{24\pi W}.
  \end{split}
\end{equation}
Here $\mathbf{p}_\pi$, $E_1$ are momentum and energy in the CMS of $\pi N$
system.

Let us express the self-energy contributions (for two channels, without
subtraction polynomials) in terms of the standard loop functions:
\begin{equation}\label{S3}
  \begin{split}
    \Sigma^1_{11}(W) &= -g_{1,\pi} \hat{\Sigma}^2_\pi g_{1,\pi} -g_{1,\eta} \hat{\Sigma}^2_\eta g_{1,\eta},  \notag \\
    \Sigma^2_{11}(W) &= -g_{1,\pi} \hat{\Sigma}^1_\pi g_{1,\pi}-g_{1,\eta} \hat{\Sigma}^1_\eta g_{1,\eta} = \Sigma^1_{11}(-W),\notag \\
    \Sigma^3_{21}(W) &= -i g_{2,\pi} \hat{\Sigma}^1_\pi g_{1,\pi}-i g_{2,\eta}\hat{\Sigma}^1_\eta g_{1,\eta}, \notag \\
    \Sigma^4_{21}(W) &= -i g_{2,\pi} \hat{\Sigma}^2_\pi g_{1,\pi}-i g_{2\,eta}\hat{\Sigma}^2_\eta g_{1,\eta} = \Sigma^3_{21}(-W), \notag \\
    \Sigma^3_{12}(W) &= \Sigma^4_{21}(W),\notag \\
    \Sigma^4_{12}(W) &= \Sigma^3_{21}(W).
  \end{split}
\end{equation}

Substituting all the necessary into \eqref{R32} we obtain the partial waves
\eqref{lab21}, \eqref{lab22}.

\end{document}